\documentclass{aa}

\usepackage{graphicx}
\usepackage{epsfig}
\usepackage{natbib}
\usepackage{txfonts}

\newcommand{\COBOLD}{{\tt CO$^5$BOLD}}
\newcommand{\LHD}{{\tt LHD}}
\newcommand{\LINFOR}{{\tt Linfor3D}}
\newcommand{\ATLAS}{{\tt ATLAS9}}
\newcommand{\teff}{\ensuremath{T_{\mathrm{eff}}}}
\newcommand{\MoH}{\ensuremath{\left[\mathrm{M}/\mathrm{H}\right]}}
\newcommand{\FeH}{\ensuremath{\left[\mathrm{Fe}/\mathrm{H}\right]}}

\begin{document}

\title{Barium abundance in  red giants of NGC 6752}
\subtitle{Non-local thermodynamic equilibrium and three-dimensional effects}

\author{
       V. Dobrovolskas
       \inst{1}
       ,
       A. Ku\v{c}inskas
       \inst{2,1}
       ,
       S. M. Andrievsky
       \inst{3,4}
       ,
       S.~A.~Korotin
       \inst{3}
       ,
       T.~V.~Mishenina
       \inst{3}
       ,
       P.~Bonifacio
       \inst{4}
       ,
       H.-G.~Ludwig
       \inst{5}
       ,
       E.~Caffau
       \inst{5,4}
       }

 \institute {Vilnius University Astronomical Observatory, M. K. \v{C}iurlionio 29, Vilnius, LT-03100, Lithuania
            \\
            \email{vidas.dobrovolskas@ff.vu.lt}
            \and
            Institute of Theoretical Physics and Astronomy, Vilnius University, Go\v{s}tauto 12, Vilnius, LT-01108, Lithuania
            \and
            Department of Astronomy and Astronomical Observatory, Odessa National University and Isaac Newton Institute of Chile Odessa branch, Shevchenko Park, 65014 Odessa, Ukraine
            \and
            GEPI, Observatoire de Paris, CNRS, Universit\'{e} Paris Diderot, Place Jules Janssen, 92190 Meudon, France
            \and
            Zentrum f\"{u}r Astronomie der Universit\"{a}t Heidelberg, Landessternwarte, K\"{o}nigstuhl 12, 69117 Heidelberg, Germany
            }

\date{Received ; accepted }

\abstract
  {}
  {We study the effects related to departures from non-local thermodynamic equilibrium (NLTE) and homogeneity in the atmospheres of red giant stars, to assess their influence on the formation of Ba~II lines. We estimate the impact of these effects on the barium abundance determinations for 20 red giants in Galactic globular cluster NGC~6752.}
  {One-dimensional (1D) local thermodynamic equilibrium (LTE) and 1D~NLTE barium abundances were derived using classical 1D \ATLAS\ stellar model atmospheres. The three-dimensional (3D) LTE abundances were obtained for 8 red giants on the lower RGB, by adjusting their 1D~LTE abundances using 3D--1D abundance corrections, i.e., the differences between the abundances obtained from the same spectral line using the 3D hydrodynamical and classical 1D stellar model atmospheres. The 3D--1D abundance corrections were obtained in a strictly differential way using the 3D hydrodynamical and classical 1D codes \COBOLD\ and \LHD. Both codes utilized identical stellar atmospheric parameters, opacities, and equation of state.}
  {The mean 1D barium-to-iron abundance ratios derived for 20 giants are $\langle{\rm [Ba/Fe]}\rangle_{\rm 1D~LTE}=0.24\pm0.05 (\rm{stat.})\pm0.08 (\rm{sys.})$ and $\langle{\rm [Ba/Fe]}\rangle_{\rm 1D~NLTE}=0.05\pm0.06 (\rm{stat.})\pm0.08 (\rm{sys.})$. The 3D--1D abundance correction obtained for 8 giants is small ($\sim+0.05$\,dex), thus leads to only minor adjustment when applied to the mean 1D~NLTE barium-to-iron abundance ratio for the 20 giants, $\langle{\rm [Ba/Fe]}\rangle_{\rm 3D+NLTE}=0.10\pm0.06(\rm{stat.})\pm0.10(\rm{sys.})$. The intrinsic abundance spread between the individual cluster stars is small and can be explained in terms of uncertainties in the abundance determinations.}
  {Deviations from LTE play an important role in the formation of barium lines in the atmospheres of red giants studied here. The role of 3D hydrodynamical effects should not be dismissed either, even if the obtained 3D--1D abundance corrections are small. This result is a consequence of subtle fine-tuning of individual contributions from horizontal temperature fluctuations and differences between the average temperature profiles in the 3D and 1D model atmospheres: owing to the comparable size and opposite sign, their contributions nearly cancel each other. This fine-tuning is characteristic of the particular set of atmospheric parameters and the element investigated, hence should not necessarily be a general property of spectral line formation in the atmospheres of red giant stars.}


\authorrunning{Dobrovolskas et al.}
\titlerunning{Barium abundance in NGC 6752}

\maketitle

\section{Introduction}

Red giants in Galactic globular clusters (GGCs) carry a wealth of important information about the chemical evolution of individual stars and their harboring populations. Owing to their intrinsic brightness, they are relatively easily accessible to high-resolution spectroscopy, thus are particularly suitable for tracing the chemical evolution histories of intermediate age and old stellar populations. Unsurprisingly, a large amount of work has been done in this direction in the past few decades \citep[for a review see, e.g.,][]{GSC04,CBG10a}, which has resulted, for example, in the discoveries of abundance anti-correlations for Na--O \citep{K94,GBB01,CBG09a}, Mg--Al \citep[see, e.g.,][]{CBG09b}, Li--Na \citep{PBM05,BPM07}, and correlation for Li--O \citep{PBM05,SBP10}.

Although the GGC stars display a scatter in their light element abundances, there is generally no spread in the abundances of iron-peak and heavier elements larger than the typical measurement errors ($\approx$0.1 dex). The only known exceptions are $\omega$ Cen \citep{SK96,NFM96} and M~54 \citep{CBG10b}, which do show noticeable star-to-star variations in the iron abundance. However, it is generally accepted that they are not genuine GGCs but instead remnants of dwarf galaxies. The first cluster where significant start-to-star variation in heavy element abundances was detected was M~15 \citep{SKS97,SPK00,OHA06,SKS11}. \citet{RS11} found that the abundances of heavy elements La, Eu, and Ho in 19 red giants of M~92 indicate that there are also significant star-to-star variations. The latter claim, however, was questioned by \citet{C11}, who found no heavy element abundance spread larger than $\sim0.07$\,dex in 12 red giants belonging to M~92. The primary formation channels of the s-process elements are the low- and intermediate-mass asymptotic giant branch (AGB) stars, thus the information about the variations in heavy element abundances may shed light on the importance of AGB stars to the chemical evolution of GGCs.

Chemical inhomogeneities involving the light elements in GGCs are the result of the products of a previous generation of stars. The nature of the stars producing these elements, or `polluters' as they are often called, remains unclear. The main contenders are rapidly rotating massive stars, that pollute the cluster through their winds \citep{decressin} or AGB stars \citep[][and references therein]{dercole}. A second order issue is whether the ``polluted'' stars are coeval and only their photospheres are polluted or they are true second-generation stars formed from the polluted material. The evidence of multiple main-sequences and sub-giant branches in GGCs \citep[see][for reviews]{piotto08,piotto09} strongly supports the latter hypothesis, although some contamination of the photospheres may still be possible.

Most of the abundance studies in GGCs have made the assumption of local thermodynamic equilibrium (LTE). Non-equilibrium effects may become especially important at low metallicity owing to the lower opacities \citep[e.g., overionization by UV photons; see, e.g.,][for more details]{A05,MGS11}. Deviations from LTE also occur because of the lower electron number density in the lower metallicity stellar atmospheres, which in turn decreases the electron collision rates with atoms and ions. Since most GGCs have metallicities that are significantly lower than solar, it is  clearly desirable to derive abundances using the non-LTE (NLTE) approach.

Nevertheless, real stars are neither stationary nor one-dimensional (1D), as assumed in the classical 1D atmosphere models that are routinely used in stellar abundance work. A step beyond these limitations can be made by using three-dimensional (3D) hydrodynamical atmosphere models that account for the three-dimensionality and non-stationarity of stellar atmospheres from first principles. Recent work has shown that significant differences may be expected between stellar abundances derived using 3D hydrodynamical and classical 1D model atmospheres (\citet{CAT07,CAN09,GBC09,RAK09,BBL10,DKL10,IKL10}; see also \citet{A05} for a review of earlier work). These differences become larger at lower metallicities and at their extremes may reach 1\,dex (!).

It is thus timely to re-analyze in a systematical and homogeneous way the abundances of various chemical elements in the GGCs, employing for this purpose state-of-the-art 3D hydrodynamical atmosphere models together with  NLTE analysis techniques. A step towards this was made in our previous work, where we derived 1D~NLTE abundances of Na, Mg, and Ba in the atmospheres of red giants belonging to GGCs M10 and M71 \citep{MKA09}. We found that in the case of the red giant N30 in M71 the 3D--1D abundance corrections for Na, Mg, and Ba, were minor and did not exceed 0.02\,dex.

In this study, we extend our previous work and derive 1D~NLTE abundances of barium in the atmospheres of 20 red giants that belong to the Galactic globular cluster NGC~6752. The analysis is done using the same techniques as in \citet{MKA09}. We also derive the 3D--1D LTE abundance corrections for the barium lines in 8 red giants and apply them to correct the 1D barium abundances for the 3D effects. Finally, we quantify the influence of both NLTE and 3D-related effects on the formation of barium lines.

The paper is organized as follows. In Sect.~2, we describe the observational material used in the abundance analysis. The procedure of barium abundance determinations is outlined in Sect.~3, where we also provide the details of the LTE/NLTE analysis and the determination of the 3D--1D abundance corrections. A discussion of our derived results is presented in Sect.~4 and the conclusions are given in Sect.~5.

\section{Observational data}

We used reduced spectra of 20 red giants in NGC~6752 available from the ESO Science Archive\footnote{http://archive.eso.org/eso/eso\_archive\_adp.html}. The high resolution ($R=60\,000$) spectral material was acquired with the UVES spectrograph at the VLT-UT2 (programme 65.L--0165(A), PI: F.~Grundahl). Spectra obtained during the three individual exposures were co-added to achieve the average signal-to-noise ratio $S/N\approx130$ at 600.0\,nm. Observations  were taken in the standard Dic 346+580 nm setting that does
not include the Ba~II 455.403 nm resonance line. The other three Ba II lines at 585.369, 614.173, and 649.691\,nm (see Table~\ref{tab:atompar}) are all found in the upper CCD of the red arm covering the range 583-680 nm. More details of the spectra acquisition and reduction procedure are provided by \citet{YGN05}. All the red giants studied in this work are located at or below the red giant branch (RGB) bump.

\section{Abundance analysis}

\subsection{Atmospheric parameters and iron abundances\label{sect:atm-par}}

Continuum normalization of the observed spectra and equivalent width ($EW$) measurements were made using the \texttt{DECH20T}\footnote{http://www.gazinur.com/DECH-software.html} software package \citep{G92}, where the $EW$s were determined using a Gaussian fit to the observed line profiles.

Stellar model atmospheres used in the abundance determinations were calculated with the Linux port version \citep{SBC04,S05} of the {\tt ATLAS9} code \citep{Kur93}, using the ODFNEW opacity distribution tables from \citet{CK03}. Models were computed using the mixing length parameter $\alpha_{\rm MLT}=1.25$ and microturbulence velocity of 1\,km~s$^{-1}$, with the overshooting option switched off. The LTE abundances were derived using the Linux port version \citep{SBC04,S05} of the Kurucz WIDTH9\footnote{{http://wwwuser.oat.ts.astro.it/castelli/sources/width9.html.}} package \citep{Kur93,Kur05,C05}.

\begin{figure}[t]
\centering
\includegraphics[width=\columnwidth]{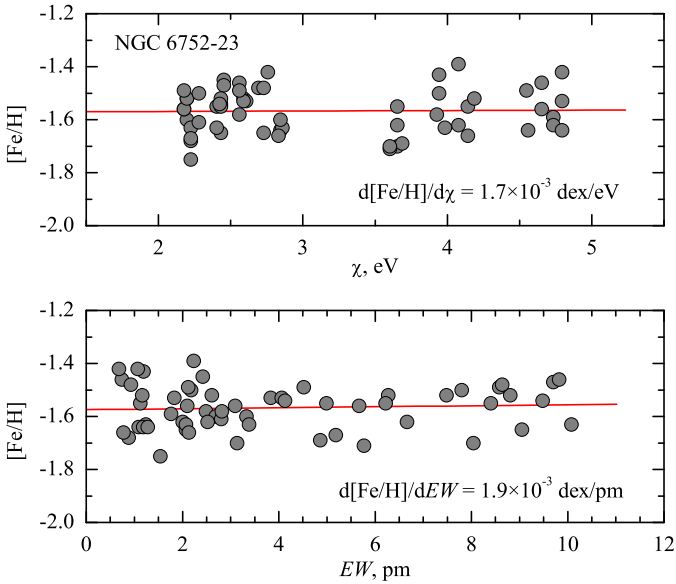}
 \caption{\FeH\ abundance ratios derived from Fe~I lines for the star NGC 6752--23, plotted versus the excitation potential (top) and line equivalent width (bottom). Linear fits to the data are shown as solid lines. Note that there is a slight correlation between the best-fit slopes in the two panels: adjusting the temperature by zeroing the slope in the upper panel slightly changes the slope in the lower panel, while changes to the microturbulence velocity influence the slope in the upper panel. This correlation does not allow us to obtain zero-valued slopes in both panels simultaneously. The adopted atmospheric parameters of this star are $\teff=4956$\,K, $\log g=2.35$, $\xi_{\rm t}=1.28$\,km~s$^{-1}$, and $\FeH=-1.57$.
 }
\label{fig:atm-par}
\end{figure}

The effective temperature, \teff, was determined under the assumption of excitation equilibrium, i.e., by requiring that the derived iron abundance
should be independent of the excitation potential, $\chi$ (Fig.~\ref{fig:atm-par}, upper panel). To obtain the value of surface gravity, $\log g$, we required that the iron abundances determined from the Fe~I and Fe~II lines would be equal. The microturbulence velocity, $\xi_{\rm t}$, was determined by requiring that Fe~I lines of different $EW$s would provide the same iron abundance (Fig.~\ref{fig:atm-par}, lower panel). The derived effective temperatures, gravities, and microturbulence velocities of individual stars agreed to within 60\,K, 0.2\,dex, and 0.16\,km~s$^{-1}$, respectively, with those determined by \citet{YGN05}.

\begin{table*}[t!]
 \begin{center}
 \caption{Target stars, their adopted atmospheric parameters, and their derived iron abundances.
 \label{tab:FeAbund}}
  \begin{tabular}{lcccrcrcccr@{}l}
  \hline
 \noalign{\smallskip}
 Star         & RA       & Dec.      & \teff, K &  $d A/d\chi$      &$\xi_{\rm t}$, km~s$^{-1}$  &  $d A/d EW$        & log \textit{g}, & $\rm [Fe/H]_{\rm I}$ &  $\rm [Fe/H]_{\rm II}$ & \multicolumn{2}{c}{A(Fe I) -- A(Fe II),} \\
              & (2000)   & (2000)    &          &  $\times 10^{-3}$, &                     &  $\times 10^{-3}$, & [cgs]           &                      &                        & \multicolumn{2}{c}{dex}                 \\
              &          &           &          &  dex/eV           &                            & dex/pm             &                 &                           &                        &                                          \\
  \hline\noalign{\smallskip}
 NGC 6752--1  & 19:10:47 & --60:00:43 & 4749     &  --2.0             & 1.37                &  --1.9             & 1.95            & --1.53                & --1.50                  & --0.&03               \\
 NGC 6752--2  & 19:11:11 & --60:00:17 & 4779     &  --3.1             & 1.35                &  --0.2             & 1.98            & --1.54                & --1.52                  & --0.&02               \\
 NGC 6752--3  & 19:11:00 & --59:56:40 & 4796     &   6.9             & 1.37                &  --0.2             & 2.03            & --1.59                & --1.60                  &  0.&01               \\
 NGC 6752--4  & 19:11:33 & --60:00:02 & 4806     &  --4.9             & 1.42                &   0.8             & 2.04            & --1.59                & --1.58                  & --0.&01               \\
 NGC 6752--6  & 19:10:34 & --59:59:55 & 4804     &  --7.6             & 1.40                &   0.3             & 1.97            & --1.58                & --1.58                  &  0.&00               \\
 NGC 6752--7  & 19:10:57 & --60:00:41 & 4829     & --11.1             & 1.39                &   0.2             & 2.10            & --1.77                & --1.75                  & --0.&02               \\
 NGC 6752--8  & 19:10:45 & --59:58:18 & 4910     &   0.9             & 1.31                &  --0.3             & 2.25            & --1.60                & --1.60                  &  0.&00               \\
 NGC 6752--9  & 19:10:26 & --59:59:05 & 4824     &   5.8             & 1.38                &  --0.9             & 2.26            & --1.61                & --1.59                  & --0.&02               \\
 NGC 6752--10 & 19:11:18 & --59:59:42 & 4836     &  --1.4             & 1.34                &   0.3             & 2.13            & --1.56                & --1.55                  & --0.&01               \\
 NGC 6752--11 & 19:10:50 & --60:02:25 & 4870     &   1.6             & 1.33                &   0.9             & 2.13            & --1.56                & --1.56                  &  0.&00               \\
 NGC 6752--12 & 19:10:20 & --60:00:30 & 4841     &   3.5             & 1.36                &   0.4             & 2.15            & --1.60                & --1.63                  &  0.&03               \\
 NGC 6752--15 & 19:10:49 & --60:01:55 & 4850     &   0.6             & 1.35                &  --1.3             & 2.19            & --1.57                & --1.56                  & --0.&01               \\
 NGC 6752--16 & 19:10:15 & --59:59:14 & 4848     &  --0.8             & 1.35                &   0.5             & 2.06            & --1.63                & --1.63                  &  0.&00               \\
 NGC 6752--19 & 19:11:23 & --59:59:40 & 4928     &  --4.0             & 1.45                &   0.1             & 2.23            & --1.64                & --1.63                  & --0.&01               \\
 NGC 6752--20 & 19:10:36 & --59:56:08 & 4929     &   2.5             & 1.25                &  --0.1             & 2.33            & --1.56                & --1.59                  &  0.&03               \\
 NGC 6752--21 & 19:11:13 & --60:02:30 & 4904     & --10.3             & 1.34                &   0.0             & 2.33            & --1.62                & --1.61                  & --0.&01               \\
 NGC 6752--23 & 19:11:12 & --59:58:29 & 4956     &   1.7             & 1.28                &   1.9             & 2.35            & --1.57                & --1.55                  & --0.&02               \\
 NGC 6752--24 & 19:10:44 & --59:59:41 & 4948     &  --2.8             & 1.19                &   1.3             & 2.28            & --1.63                & --1.63                  &  0.&00               \\
 NGC 6752--29 & 19:10:17 & --60:01:00 & 4900     &   1.9             & 1.31                &  --2.0             & 2.24            & --1.68                & --1.63                  & --0.&05               \\
 NGC 6752--30 & 19:10:39 & --59:59:47 & 4943     &  --4.5             & 1.18                &  --0.3             & 2.42            & --1.64                & --1.62                  & --0.&02               \\
 \hline
 mean         &          &           &          &                   &                     &                   &                 & --1.60                & --1.60                  &   &                  \\
 $\sigma$     &          &           &          &                   &                     &                   &                 &  0.05                &  0.05                  &     &                \\
 \hline
 \end{tabular}
 \end{center}
 Note: Fe~II abundances were adjusted to match the abundances determined from Fe~I lines, in order to estimate the surface gravities of the target stars. The difference between the corresponding abundance ratios, $\FeH_{\rm I}$, and $\FeH_{\rm II}$, is thus only indicative of the goodness of the gravity estimates.
\end{table*}

The LTE iron abundances for all stars in our sample were derived using 50-60 neutral iron lines (Table~\ref{tab:FeLines}; note that the iron abundance derived from the ionized lines was required to match that of neutral iron, i.e., to obtain the estimate of surface gravity, thus it is not an independent iron abundance measurement). To minimize the impact of NLTE effects on the iron abundance determinations, we avoided neutral iron lines with the excitation potential $\chi < 2.0$ eV. Oscillator strengths and damping constants for all iron lines were retrieved from the VALD database \citep{KRP00}. The obtained iron abundances are provided in Table~\ref{tab:FeAbund}. The contents of the table are as follows: the star identification and its coordinates are given in Cols.~1--3, effective temperatures and iron abundance derivatives relative to the excitation potential are in columns~4 and 5, respectively, the adopted microturbulence velocity and iron abundance derivative relative to the equivalent width are in columns~6 and 7, respectively, the adopted values of $\log g$ are in column~8, iron abundances obtained from Fe~I and Fe~II lines are in columns~9 and 10, respectively, and the difference between them is in column~11.

The mean iron abundance obtained for the 20 stars is $\langle\FeH\rangle=-1.60\pm0.05$, which is in excellent agreement with  $\FeH=-1.62\pm0.02$ obtained by \citet{YGN05}.

\begin{table}[tb]
 \begin{center}
\caption{Atomic parameters of the barium lines used in this work.
\label{tab:atompar}}
  \begin{tabular}{lcccccc}
  \hline
\noalign{\smallskip}
 $\lambda$$^{\mathrm{a}}$, nm  & $\chi$$^{\mathrm{a}}$, eV & log \textit{gf}$^{\mathrm{a}}$ & log $\gamma_{rad}$$^{\mathrm{b}}$ & log $\frac{\gamma_4}{N_e}$$^{\mathrm{c}}$ & log $\frac{\gamma_6}{N_H}$$^{\mathrm{d}}$ \\
  \hline\noalign{\smallskip}
 585.3688       & 0.604     & --1.000          & 8.20    & --5.460 & --7.190 \\
 614.1730       & 0.704     & --0.076          & 8.20    & --5.480 & --7.470 \\
 649.6910       & 0.604     & --0.377          & 8.10    & --5.480 & --7.470 \\
  \hline
  \end{tabular}
  \end{center}
  \begin{list}{}{}
\item[$^{\mathrm{a}}$] \citet{WM80}; $^{\mathrm{b}}$ natural broadening constant, from \citet{MB96}; $^{\mathrm{c}}$ Stark broadening constant, from \citet{KRP00}; $^{\mathrm{d}}$ van der Waals broadening constant, from \citet{KMG11}
\end{list}
\end{table}

\begin{table*}[tbh]
 \begin{center}
\caption{Measured equivalent widths of the barium lines and the derived barium abundances in individual red giants in NGC~6752.
\label{tab:EWba}}
  \begin{tabular}{lccccccr@{}l}
  \hline
  \noalign{\smallskip}
Star         & \textit{EW} (585.4nm), & \textit{EW} (614.2nm), &  \textit{EW} (649.7nm), & A(Ba)$_\mathrm{1D~LTE}$, & A(Ba)$_\mathrm{1D~NLTE}$,  & [Ba/Fe]$_\mathrm{1D~LTE}$, & \multicolumn{2}{c}{[Ba/Fe]$_\mathrm{1D~NLTE}$,} \\
             &     pm        &     pm        &      pm        &    dex   &   dex    &    dex      & &     dex        \\
 \hline\noalign{\smallskip}
NGC~6752--1   &  8.21  &  12.54  &  12.87  &  0.92 &  0.63 &  0.28  &  --0.&01 \\
NGC~6752--2   &  8.43  &  12.45  &  12.08  &  0.91 &  0.65 &  0.28  &    0.&02 \\
NGC~6752--3   &  7.21  &  11.84  &  11.65  &  0.75 &  0.50 &  0.17  &  --0.&08 \\
NGC~6752--4   &  8.25  &  12.07  &  12.26  &  0.86 &  0.68 &  0.28  &    0.&10 \\
NGC~6752--6   &  8.23  &  11.96  &  12.80  &  0.87 &  0.60 &  0.28  &    0.&01 \\
NGC~6752--7   &  7.13  &  10.72  &  11.32  &  0.60 &  0.44 &  0.20  &    0.&04 \\
NGC~6752--8   &  7.01  &  11.29  &  11.37  &  0.84 &  0.72 &  0.27  &    0.&15 \\
NGC~6752--9   &  7.50  &  11.65  &  12.02  &  0.85 &  0.70 &  0.29  &    0.&14 \\
NGC~6752--10  &  7.57  &  11.58  &  12.16  &  0.87 &  0.67 &  0.26  &    0.&06 \\
NGC~6752--11  &  7.59  &  11.62  &  11.13  &  0.82 &  0.61 &  0.21  &    0.&00 \\
NGC~6752--12  &  7.01  &  11.44  &  11.34  &  0.76 &  0.61 &  0.19  &    0.&04 \\
NGC~6752--15  &  7.13  &  11.24  &  11.35  &  0.78 &  0.60 &  0.18  &    0.&00 \\
NGC~6752--16  &  7.36  &  11.51  &  11.59  &  0.80 &  0.58 &  0.26  &    0.&04 \\
NGC~6752--19  &  6.84  &  10.76  &  11.03  &  0.69 &  0.57 &  0.16  &    0.&04 \\
NGC~6752--20  &  6.77  &  11.16  &  10.81  &  0.87 &  0.75 &  0.26  &    0.&14 \\
NGC~6752--21  &  7.29  &  10.74  &  11.09  &  0.82 &  0.64 &  0.27  &    0.&09 \\
NGC~6752--23  &  7.17  &  10.71  &  11.21  &  0.88 &  0.72 &  0.28  &    0.&12 \\
NGC~6752--24  &  6.03  &   9.79  &   9.57  &  0.68 &  0.53 &  0.14  &  --0.&01 \\
NGC~6752--29  &  6.35  &  10.06  &  10.52  &  0.67 &  0.51 &  0.18  &    0.&02 \\
NGC~6752--30  &  6.39  &   9.92  &  10.57  &  0.82 &  0.66 &  0.29  &    0.&13 \\
  \hline
mean         &        &         &         &  0.80 &  0.62 &  0.24  &    0.&05 \\
$\sigma$     &        &         &         &  0.09 &  0.08 &  0.05  &    0.&06 \\
  \hline
  \end{tabular}
  \end{center}
\end{table*}

\subsection{One-dimensional LTE abundances of barium\label{sect:lte-abund}}

One-dimensional (1D) LTE barium abundances were derived from the three Ba~II lines centered at 585.3688\,{nm}, 614.1730\,{nm}, and 649.6910\,{nm}. Damping constants and other atomic parameters of the three barium lines are provided in Table~\ref{tab:atompar}. The line equivalent widths were measured with the \texttt{DECH20T} software (Table~\ref{tab:EWba}, columns~2--4). Hyperfine splitting of the 649.6910\,{nm} line was not taken into account in the 1D~LTE analysis. The derived barium abundances and barium-to-iron abundance ratios are given in Table~\ref{tab:EWba}, columns~5 and 7, respectively.

We note that the barium line at 614.1730\,{nm} is blended with a neutral iron line located at 614.1713\,{nm}. To estimate how this affects the accuracy of the abundance determination, we synthesized the barium 614.1730\,{nm} line with and without the blending iron line, for all stars in our sample. The comparison of the equivalent widths of blended and non-blended lines reveals that the contribution of the iron blend never exceeds $\sim2.4$\,\%, or $\leq0.05$\,dex in terms of the barium abundance. The contribution of the iron blend to the $EW$ of the 614.1730\,{nm} line was thus taken into account by reducing the measured equivalent widths of this barium line by 2.4\,\% for all stars. We would like to point out, however, that in the 1D~NLTE analysis the barium abundances were derived by fitting the synthetic spectrum to the observed line profile, thus the influence of the iron blend at 614.1713\,{nm} was properly taken into account.

Assessment of the abundance sensitivity on the atmospheric parameters yields the following results:

\begin{itemize}
\item
Change in the effective temperature by ${\pm}80$\,K leads to a change in the barium abundance measured from the three Ba~II lines by $\mp0.03$\,dex;
\item
Change in the surface gravity by $\pm0.1$\,dex changes the barium abundance by $\mp0.02$\,dex;
\item
Change in the microturbulence velocity, $\xi_{\rm t}$, by $\pm0.1$\,km~s$^{-1}$ changes the barium abundance by $\mp0.07$\,dex.
\end{itemize}

Since barium lines in the target stars are strong and situated in the saturated part of the curve of growth, it is unsurprising that the uncertainty in the microturbulence velocity is the largest contributor to the uncertainty in the derived barium abundance. The total contribution from the individual uncertainties in \teff, $\log g$, and $\xi_{\rm t}$ leads to the systematic uncertainty in the barium abundance determinations of $\sim0.08$\,dex. We note, however, that the latter number does not account for the uncertainty in the equivalent width determination and thus only provides the lower limit to the systematic uncertainty (e.g., 5 percent in the equivalent width determination leads to the barium abundance uncertainty of $\sim0.1$\,dex).

The obtained mean 1D~LTE barium abundance for the sample of 20 stars in NGC~6752 is ${\rm \langle A(Ba)\rangle}_{\rm 1D~LTE}=0.80\pm0.09\pm0.08$ and the barium-to-iron ratio is ${\rm \langle[Ba/Fe]\rangle}_{\rm 1D~LTE}=0.24\pm0.05\pm0.08$. In both cases, the first error is a square root of the variance calculated for the ensemble of individual abundance estimates of 20 stars. The second error is the systematic uncertainty in the atmospheric parameter determination. The difference between the individual barium abundances derived in a given star using the three barium lines is always below $\sim0.1$\,dex.

\begin{figure*}
\centering
\includegraphics[width=16cm]{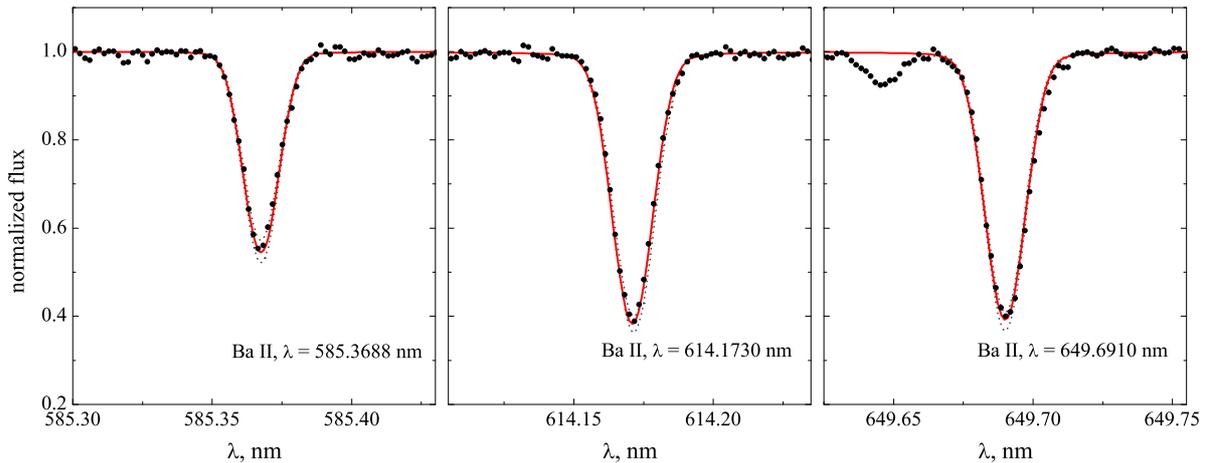}
 \caption{Fit of the synthetic NLTE profiles of the Ba~II lines (solid line) to the observed spectrum of the red giant NGC6752--23 (dots). Synthetic line profiles corresponding to the abundances $0.1$\,dex higher/lower than the best-fit value are shown as dashed lines.}
\label{fig:line-prof}
\end{figure*}

\subsection{One-dimensional NLTE abundances of barium\label{sect:nlte-abund}}

The one dimensional (1D) NLTE abundances of barium were determined using the version of the 1D~NLTE spectral synthesis code {\tt MULTI} \citep{carlsson} modified by \citet{KAL99}. The model atom of barium used in the NLTE spectral synthesis calculations was taken from \citet{ASK09}. To summarize briefly, it consisted of 31 levels of Ba\,I, 101 levels of Ba\,II ($n<50$) and the ground level of Ba\,III. In total, 91 bound-bound transitions were taken into account between the first 28 levels of Ba\,II ($n<12$, $l<5$). Fine structure was taken into account for the levels 5d$^{2}$D and 6p$^{2}$P$^{0}$, according to the prescription given in \citet{ASK09}. We also accounted for the hyperfine splitting of the barium 649.6910\,{nm} line. Isotopic splitting of the barium lines was not taken into account. Owing to the low ionization potential of neutral barium ($\sim5.2$\,eV), Ba~II is the dominant ionization stage in the line-forming regions of investigated stars, with $n({\rm Ba~I})/n({\rm Ba~II})\lesssim10^{-4}$. It is therefore safe to assume that none of the Ba~I transitions may noticeably change the level populations of Ba II \citep[cf.][]{MGB99}. Further details about the barium model atom, the assumptions used, and implications involved can be found in \citet{ASK09} and \citet{KMG11}.

The solar abundances of iron and barium were assumed to be $\log A({\rm Fe})_{\odot}=7.50$ and $\log A({\rm Ba})_{\odot}=2.17$ respectively, on the scale where $\log A({\rm H})_{\odot}=12$. These abundances were determined using the Kurucz Solar Flux Atlas \citep{KFB84} and the same NLTE approach as applied in this study.

A typical fit of the synthetic line profiles to the observed spectrum is shown in Fig.~\ref{fig:line-prof}, where we plot synthetic and observed profiles of all three barium lines used in the analysis. The elemental abundances and barium-to-iron abundance ratios derived for the individual cluster giants are provided in Table~\ref{tab:EWba} (columns~6 and 8, respectively).

The mean derived 1D~NLTE barium-to-iron ratio for the 20 cluster red giants is ${\rm \langle[Ba/Fe]\rangle}_{\rm 1D~NLTE}=0.05\pm0.06\pm0.08$. The first error is the square root of the variance in ${\rm \langle[Ba/Fe]\rangle}_{\rm 1D~NLTE}$ estimates obtained for the ensemble of 20 stars, thus measures the star-to-star variation in the barium-to-iron ratio. The second error is the systematic uncertainty resulting from the stellar parameter determination (see Section 3.1). The individual line-to-line barium abundance scatter was always significantly smaller than 0.1\,dex.

We find that barium lines generally appear stronger in NLTE than in LTE, which leads to lower NLTE barium abundances. This is in accord with the results obtained by \citet{SH06} for the metallicity of NGC~6752, and similar to the trends obtained for cool dwarfs by \citet{MGB99}. The NLTE--LTE abundance corrections for the three individual barium lines are always very similar, with the differences being within a few hundredths of a dex.

\subsection{3D--1D barium abundance corrections}

We have used the \COBOLD\ 3D hydrodynamical \citep{FSL12} and \LHD\ 1D hydrostatic \citep{CL07} stellar atmosphere models  to investigate how strongly the formation of barium lines may be affected by convective motions in the stellar atmosphere. The \COBOLD\ code solves the 3D equations of radiation hydrodynamics under the assumption of LTE. The model assumes a cartesian coordinate grid. For a detailed description of the \COBOLD\ code and its applications, we refer to \citet{FSL12}.

Since we did not have \COBOLD\ models available for the entire atmospheric parameter range covered by the red giants in NGC~6752, we estimated the importance of 3D hydrodynamical effects only for stars on the lower RGB. For this purpose, we used a set of 3D hydrodynamical \COBOLD\ models with $\teff=5000$\,K and $\log g=2.5$, at four different metallicities, \MoH = 0.0, --1.0, --2.0, and --3.0\footnote{The models at four metallicities were needed to interpolate the 3D--1D abundance corrections at the metallicity of NGC~6752.}. Allowing for the error margins of $\sim100$\,K in the effective temperature and $\sim0.25$\,dex in gravity, we assumed that the effective temperature and gravity of this model set is representative of the atmospheric parameters of the stars NGC~6752--08, and NGC~6752--19 to NGC6752--30 (8 objects, see Table~\ref{tab:FeAbund}). For these stars, the extreme deviations from the parameters of the 3D model are $\Delta\teff\sim110$\,K and $\Delta\log g\sim0.26$. These differences would only have a marginal effect on the uncertainty in the abundance estimates, i.e., the systematic uncertainty for the 3D barium abundance derivations would only increase from $\pm0.08$\,dex quoted in Sect.~\ref{sect:lte-abund} to $\pm0.10$\,dex.

\begin{table}[tb]
\caption{Parameters of the 3D hydrodynamical \COBOLD\ atmosphere models used in this work.\label{tab:3Dmodels}}   
\label{table1}             
\centering                 
\begin{tabular}{c c c c c}     
\hline                     
  $T_{\rm eff}$, K & $\log g$ & [M/H] & Grid dimension, Mm & resolution\\    
                   &          &       & x $\times$ y $\times$ z & x $\times$ y $\times$ z\\
\hline                     
 4970 & 2.5 &   0    & 573\(\times\)573\(\times\)243 & 160\(\times\)160\(\times\)200\\    
 4990 & 2.5 & \(-1\) & 573\(\times\)573\(\times\)245 & 160\(\times\)160\(\times\)200\\
 5020 & 2.5 & \(-2\) & 584\(\times\)584\(\times\)245 & 160\(\times\)160\(\times\)200\\
 5020 & 2.5 & \(-3\) & 573\(\times\)573\(\times\)245 & 160\(\times\)160\(\times\)200\\
\hline                     
\end{tabular}
\end{table}

\begin{figure}[tbh]
\centering
\includegraphics[width=\columnwidth]{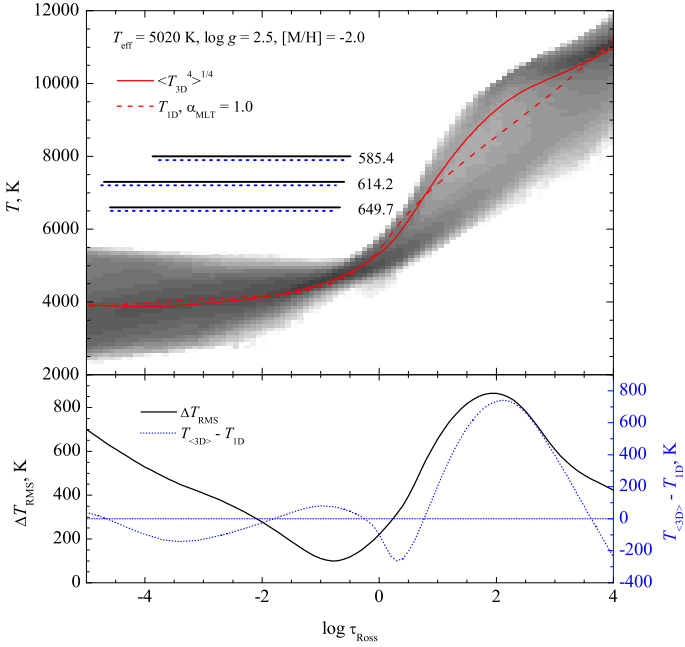}
 \caption{Top panel: Temperature stratification in a single snapshot of the 3D hydrodynamical \COBOLD\ model at the metallicity \MoH = --2. Gray shaded area shows the temperature probability density on a logarithmic scale, with darker shades meaning a higher probability of finding a particular temperature value in the 3D model simulation box. The solid red line shows mean temperature stratification of the 3D model and the dashed red line is the 1D \LHD\ model temperature stratification. Horizontal bars show the optical depth intervals where 90\% of the line equivalent width is formed: black bars correspond to the 3D model while blue dashed correspond to the 1D. Numbers next to the bars designate the wavelength of the given Ba~II line in nm. Bottom panel: RMS value of horizontal temperature fluctuations in the 3D model (black line) and temperature difference between the mean 3D and 1D models (blue dashed line).}
 \label{fig:temp-strat}
\end{figure}

The 3D hydrodynamical models were taken from the CIFIST 3D model atmosphere grid \citep{LCS09}. The model parameters are summarized in Table~\ref{tab:3Dmodels}. The physical size of the 3D model box was chosen so that it would accommodate at least ten convective cells in the horizontal plane. Monochromatic opacities used in the model calculations were grouped into five opacity bins for $\MoH=0.0$ and six opacity bins for $\MoH=-1.0,-2.0,-3.0$ models. The 1D \LHD\ models were calculated for the same set of atmospheric parameters using the same equation of state, opacities, and chemical composition as in the case of the 3D hydrodynamical models.

To illustrate the differences between the 3D hydrodynamical and 1D classical stellar model atmospheres, we show their temperature stratifications at the metallicity of $\MoH=-2.0$, which is the closest to that of NGC~6752 (Fig.~\ref{fig:temp-strat}, upper panel). In the same figure, we also indicate the typical formation depths of the three barium lines. It is obvious that at these depths, the temperature of the 3D hydrodynamical model fluctuates very strongly, especially in the outer atmosphere, as indicated by the RMS value of horizontal temperature fluctuations ($\Delta T_{\rm{RMS}} = \sqrt{\langle(T - T_0)^2\rangle_{x,y,t}}$, where $T_0$ is the temporal and horizontal temperature average obtained on surfaces of equal optical depth). As we see below, differences in the atmospheric structures lead to differences in the line formation properties and henceforth to differences in barium abundances obtained with the 3D hydrodynamical and 1D classical model atmospheres.

Twenty 3D snapshots (i.e., 3D model structures at different instants in time) were selected to calculate the Ba~II line profiles. The snapshots were chosen in such a way that the statistical properties of the snapshot sample (average effective temperature and its r.m.s value, mean velocity at the optical depth unity, etc.) would match as close as possible those of the entire ensemble of the 3D model run. The 3D line spectral synthesis was performed for each individual snapshot and the resulting line profiles were averaged to yield the final 3D spectral line profile.

\begin{table*}[tb]
 \begin{center}
 \caption{The 3D~LTE barium abundances of red giants in NGC~6752.
 \label{tab:3D_Ba_Abn}}
  \begin{tabular}{lr@{ }rr@{ }rr@{ }rrccc r@{.}l r@{.}l}
  \hline
  \noalign{\smallskip}
Star & \multicolumn{6}{c}{$\Delta_{\rm 3D-1D}^{\rm line}$, $\Delta_{\rm 3D-\langle3D\rangle}^{\rm line}$, dex}  & $\Delta_{\rm 3D-1D}$   & $\xi_{\rm t}$  & A(Ba)  &  [Ba/Fe] & \multicolumn{2}{c}{[Ba/Fe]} & \multicolumn{2}{c}{[Ba/Fe]}\\
  \noalign{\smallskip}
            & \multicolumn{2}{c}{BaII5854} & \multicolumn{2}{c}{BaII6142} & \multicolumn{2}{c}{BaII6497} &  dex  & km~s$^{-1}$   & 3D~LTE& 3D~LTE  & \multicolumn{2}{c}{1D~NLTE}  & \multicolumn{2}{c}{3D+NLTE} \\
  \hline\noalign{\smallskip}
NGC 6752--08 &  0.05  & 0.00   &   0.08 & --0.01 &   0.06 & --0.03 &  0.06 &  1.31  &  0.90 &    0.33 &    0&15 &    0&21 \\
NGC 6752--19 &  0.13  & 0.08   &   0.18 &   0.11 &   0.15 &   0.07 &  0.15 &  1.45  &  0.84 &    0.31 &    0&04 &    0&19 \\
NGC 6752--20 &  0.01  & --0.04 &   0.04 & --0.05 &   0.02 & --0.07 &  0.02 &  1.25  &  0.89 &    0.28 &    0&14 &    0&16 \\
NGC 6752--21 &  0.07  & 0.02   &   0.10 &   0.02 &   0.08 & --0.01 &  0.08 &  1.34  &  0.90 &    0.35 &    0&09 &    0&17 \\
NGC 6752--23 &  0.02  & --0.03 &   0.06 & --0.02 &   0.04 & --0.05 &  0.04 &  1.28  &  0.92 &    0.32 &    0&12 &    0&16 \\
NGC 6752--24 & --0.02 & --0.07 & --0.01 & --0.08 & --0.03 & --0.10 & --0.02&  1.19  &  0.66 &    0.12 &  --0&01 &  --0&03 \\
NGC 6752--29 &  0.05  & 0.00   &   0.08 &   0.01 &   0.06 & --0.02 &  0.06 &  1.31  &  0.73 &    0.24 &    0&02 &    0&08 \\
NGC 6752--30 & --0.03 & --0.08 & --0.01 & --0.09 & --0.02 & --0.11 & --0.02&  1.18  &  0.80 &    0.27 &    0&13 &    0&11 \\
  \hline
            &  &  &  &  &  &  &       &   mean:&  0.83 &    0.28 &    0&08 &    0&13 \\
            &  &  &  &  &  &  &       &  sigma:&  0.09 &    0.07 &    0&06 &    0&08 \\
  \hline
  \end{tabular}
  \end{center}
\end{table*}

The influence of convection on the spectral line formation was estimated by means of 3D--1D abundance corrections. The 3D--1D abundance correction is defined as the difference between the abundance $A(Y)$ derived for a given element $Y$ from the same observed spectral line using the 3D hydrodynamical and classical 1D model atmospheres, i.e., $\Delta_{\rm 3D-1D}=A(Y)_{\rm 3D}-A(Y)_{\rm 1D}$ \citep{CLS11}. This abundance correction can be separated into two constituents: (a) correction owing to the horizontal temperature inhomogeneities in the 3D model, $\Delta_{\rm
3D-\langle3D\rangle}=A(Y)_{\rm 3D}-A(Y)_{\rm \langle3D\rangle}$, and (b) correction owing to the differences between the temperature profiles of the average ${\rm \langle3D\rangle}$ and 1D models, $\Delta_{\rm \langle3D\rangle-1D}=A(Y)_{\rm \langle3D\rangle}-A(Y)_{\rm 1D}$. Abundances corresponding to the subscript ${\rm \langle3D\rangle}$ were derived using the average 3D models, which were obtained by horizontally averaging 3D model snapshots on surfaces of equal optical depth. Spectral line profiles were calculated for each average ${\rm \langle3D\rangle}$ structure corresponding to individual 3D model snapshots. These line profiles were averaged to yield the final ${\rm \langle3D\rangle}$ profile, which was used to derive the $\Delta_{\rm \langle3D\rangle-1D}$ abundance corrections. The full abundance correction was then $\Delta_{\rm 3D-1D}\equiv\Delta_{\rm 3D-\langle3D\rangle}+\Delta_{\rm \langle3D\rangle-1D}$. Spectral line synthesis for all three models, i.e., 3D, ${\rm \langle3D\rangle}$, and 1D, was made using the \LINFOR\ code\footnote{http://www.aip.de/\textasciitilde mst/Linfor3D/linfor\_3D\_manual.pdf}.

The barium lines in the target stars are strong (cf. Table~\ref{tab:EWba}) and thus the derived 3D--1D abundance corrections are sensitive to the  microturbulence velocity, $\xi_{\rm t}$, of the comparison 1D model. The 3D--1D abundance corrections were therefore calculated using the equivalent widths and microturbulence velocities of the target stars derived in Sect.~\ref{sect:atm-par} and \ref{sect:lte-abund}. Furthermore, cubic interpolation was used to interpolate between the 3D--1D abundance corrections derived at four different metallicities to obtain its value at the metallicity of the cluster, $\FeH=-1.6$. The cubic interpolation between the four values of metallicities was chosen because of the nonlinear dependence of the 3D--1D abundance corrections on metallicity. The results are provided in Table~\ref{tab:3D_Ba_Abn}, which contains the $\Delta_{\rm 3D-1D}$ and $\Delta_{\rm 3D-\langle3D\rangle}$ abundance corrections for the three individual Ba~II lines (columns~2--4), the 3D--1D abundance correction for each star (i.e., averaged over the three barium lines, column~5), the microturbulence velocity used with the 1D comparison model (column~6, from Sect.~\ref{fig:atm-par}), the 3D~LTE barium abundances (column~7), the 3D~LTE barium-to-iron ratio (column~8), and finally both the 1D~NLTE barium-to-iron ratio before (column~9, from Sect.~\ref{sect:nlte-abund}) and after correction for the 3D effects (column~10).

Abundance corrections are sensitive to the choice of the 1D microturbulence velocity and line strength, therefore stars with very similar atmospheric parameters may have different abundance corrections. This is, for example, the case for NGC~6752-19 and NGC~6752-30. These two stars have largest and smallest microturbulence velocities in the entire sample, respectively, and NGC6752-19 has slightly stronger barium lines than NGC~6752-30 (Table~\ref{tab:EWba}). This leads to noticeably different abundance corrections, despite both stars having very similar effective temperatures and gravities (Table~\ref{tab:3D_Ba_Abn}).

\section{Discussion}

\subsection{One-dimensional LTE and NLTE barium abundances in NGC~6752}

Generally, the mean 1D~LTE barium-to-iron abundance ratio obtained in this work, ${\rm \langle [Ba/Fe]\rangle}_{\rm1D~LTE}=0.24\pm0.05$\,(random) $\pm$\,0.08\,(systematic), agrees well with the 1D~LTE abundance ratios derived for this cluster by other authors. For example, \citet{JFB04} derived ${\rm \langle[Ba/Fe]\rangle}=0.18\pm0.11$ from 9 main-sequence and 9 subgiant stars in NGC~6752. We note that the mean barium-to-iron ratio of \citet{JFB04} based on the measurements of only subgiant stars is ${\rm \langle [Ba/Fe]\rangle}=0.25\pm0.08$, i.e., in this case the agreement with our LTE estimate is even better. The barium-to-iron ratios obtained by \citet{NDC95} and \citet{YGN05} are lower, ${\rm \langle [Ba/Fe]\rangle}=0.00\pm0.13$ and ${\rm \langle [Ba/Fe]\rangle}=-0.06\pm0.13$, respectively.

The disagreement between the results of \citet{YGN05} and those obtained here is somewhat concerning, especially since both studies were based on the same set of UVES spectra, while the atmospheric parameters and iron abundances of individual stars employed by us and \citet{YGN05} agree very well (Sect.~\ref{sect:atm-par}). Moreover, the comparison of the equivalent width measurements obtained by us and \citet{YGN05} also shows good agreement. One would thus also expect good agreement in the derived barium abundances -- which is unfortunately not the case. We therefore felt it was important to look into the possible causes of this discrepancy.

To this end, we first obtained the 1D~LTE barium abundances using the {\tt MULTI} code. This independent abundance estimate was made using the same procedure as for the 1D~NLTE abundance derivations, i.e., by fitting the observed and synthetic line profiles of the three Ba~II lines, with the difference that in this case the line profile calculations performed with \texttt{MULTI} were done under the assumption of LTE. The mean barium-to-iron abundance ratio obtained in this way, ${\rm \langle[Ba/Fe]\rangle}=0.22\pm0.06\pm0.08$, agrees well with the value derived in Section~3.2 (${\rm \langle[Ba/Fe]\rangle}=0.24\pm0.05\pm0.08$).

In their abundance determinations, \citet{YGN05} used an older version of the {\tt ATLAS} models \citep{Kur93}. The differences between these {\tt ATLAS} models and those used in our work is that (a) different opacity tables were used in the two cases \citep[i.e., ODFNEW from][with our models]{CK03}, and (b) the {\tt ATLAS} models of \citet{Kur93} were calculated with the overshooting parameter switched on, while in our case the overshooting was switched off. To check the influence of these differences on the abundance derivations, we obtained the 1D~LTE barium abundance using the older atmosphere models of \citet{Kur93}, with the atmospheric parameters and iron abundances derived in Sect.~\ref{sect:atm-par}. In this case, the mean derived barium-to-iron abundance ratio was ${\rm \langle[Ba/Fe]\rangle}_{\rm 1D~LTE}=0.23\pm0.05\pm0.08$, i.e., the effect of differences in the model atmospheres was only $\sim0.01$\,dex. The change in the barium abundances owing to differences in the atomic parameters (line broadening constants, oscillator strengths) used in the two studies was more significant, i.e., the abundances derived by us using the atomic parameters of \citet{YGN05} were $\sim0.1$\,dex lower. However, this still leaves a rather large discrepancy, $\sim0.15$\,dex, between the barium-to-iron ratios obtained by us and \citet{YGN05}, for which we unfortunately cannot find a plausible explanation.

As in the case of the 1D~LTE abundances, the extent of the star-to-star variations in the derived 1D~NLTE barium-to-iron ratio, ${\rm \langle [Ba/Fe]\rangle}_{\rm1D\,NLTE}=0.05\pm0.06\pm0.08$, is small and can be fully explained by the uncertainties in the abundance determination. The 1D~NLTE barium-to-iron ratio derived here is similar to the value ${\rm \langle [Ba/Fe]\rangle}_{\rm 1D\,NLTE}=0.09\pm0.20$ obtained for two red giants in M10 by \citet{MKA09}. The elemental ratios obtained in the two studies are thus very similar, although one should keep in mind that the estimate of \citet{MKA09} is based on only two stars. The metallicities of the two clusters are very similar too, ${\rm [Fe/H]}=-1.56$ in the case of M10 \citep[][2010 edition]{H96} and ${\rm [Fe/H]}=-1.60$ for NGC~6752 (Sect.~\ref{sect:atm-par}). Galactic field stars typically show no pronounced dependence of [Ba/Fe] on metallicity, although the scatter at any given metallicity is large \citep{sneden}. One may therefore conclude that, taken into account the high [Ba/Fe] spread in field stars, the ${\rm \langle[Ba/Fe]\rangle}$ ratio derived here is comparable to those seen in Galactic field stars and other globular clusters of similar metallicity.

\subsection{The 3D-corrected 1D NLTE barium abundance in NGC~6752}

The 3D--1D barium abundance corrections obtained for the eight stars in NGC~6752 (see Section 3.4 above) provide a hint of the net extent to which the 3D hydrodynamical effects may influence spectral line formation (and thus, the abundance determinations) in their atmospheres (Table~\ref{tab:3D_Ba_Abn}). In the case of all red giants investigated, the corrections are small, --0.03 to +0.15\,dex, and the mean abundance correction for the eight stars is $\Delta_{\rm 3D-1D}=0.05$. We note though that the individual contributions to the abundance correction, $\Delta_{\rm 3D-\langle3D\rangle}$ and $\Delta_{\rm \langle3D\rangle-1D}$, are substantial ($\sim\pm0.1$\,dex) but often because of their opposite sign nearly cancel and thus the resulting abundance correction is significantly smaller (Table~\ref{tab:3D_Ba_Abn}). This clearly indicates that the role of convection-related effects on the spectral line formation in these red giants cannot be neglected, even if the final 3D--1D abundance correction,  $\Delta_{\rm 3D-1D}$, is seemingly very small.

The mean 3D~LTE barium-to-iron abundance ratio obtained for the eight red giants is ${\rm \langle [Ba/Fe]\rangle}_{\rm3D~LTE}=0.28\pm0.07\pm0.10$. The 3D~LTE barium abundance measurements made for a given star from the three barium lines always agree to within $\approx0.03$\,dex. In the case of all twenty giants studied here, the mean 1D~NLTE barium-to-iron ratio corrected for the 3D-related effects is ${\rm \langle [Ba/Fe]\rangle}_{\rm3D+NLTE}=0.10\pm0.08\pm0.10$ and therefore is only slightly different from the 1D~NLTE value obtained in Sect.~\ref{sect:nlte-abund}. However, the positive sign of the 3D--1D abundance differences indicates that in the spectra of red giants in NGC~6752 the three studied Ba~II lines will be \textit{weaker} in 3D than in 1D, in contrast to what is generally seen in red giants at this metallicity \citep[cf.][]{CAT07,DKL10}.

For the Ba~II lines, the 3D--1D abundance corrections are sensitive to the choice of microturbulence velocities in the 1D models: an increase in the microturbulence velocity by 0.10\,km~s$^{-1}$ leads to an increase of 0.07\,dex in the 3D--1D abundance correction.
At the same time, the 1D abundance itself {\em decreases} by roughly
the same amount. The result is that although the 3D correction is
sensitive to microturbulence, the 3D corrected abundance is not.

\section{Conclusions}

We have derived the 1D~LTE and 1D~NLTE abundances of barium for 20 red giant stars in the globular cluster NGC~6752. The mean barium-to-iron abundance ratios are ${\rm \langle [Ba/Fe]\rangle_{\rm 1D~LTE}}=0.24\pm0.05\pm0.08$ and ${\rm \langle [Ba/Fe]\rangle_{\rm 1D~NLTE}}=0.05\pm0.06\pm0.08$ (the first error measures the star-to-star variation in the abundance ratio and the second is the systematic uncertainty in the atmospheric parameter determination, see Sect.~\ref{sect:atm-par}). Individual barium-to-iron abundance ratios show little star-to-star variation, which leads us to conclude that there is no intrinsic barium abundance spread in the RGB stars at or slightly below the RGB bump in NGC~6752. This conclusion is in line with the results obtained in other studies, for stars in both this and other GGCs \citep{NDC95,JFB04,YGN05}.

The derived 1D~NLTE barium-to-iron abundance ratio is comparable to the one observed in Galactic halo stars of the same metallicity \citep{sneden}. It is also similar to the mean barium-to-iron abundance ratio obtained by \citet{MKA09} for 2 red giants in the Galactic globular cluster M10. We therefore conclude that the barium-to-iron abundance ratios obtained here generally agree with those seen in the oldest Galactic populations and are not very different from those observed in halo stars.

We have also obtained 3D~LTE barium abundances for 8 red giants on the lower RGB in NGC~6752. The mean 3D~LTE barium abundance, ${\rm \langle [Ba/Fe]\rangle}_{\rm3D~LTE}=0.28\pm0.07\pm0.10$, is only $0.05$\,dex higher than that obtained for these stars in 1D~LTE. This small 3D--1D correction leads to very minor adjustment of the mean 1D~NLTE barium-to-iron ratio for the 20 investigated giants, ${\rm \langle [Ba/Fe]\rangle}_{\rm3D+NLTE}=0.10\pm0.08\pm0.10$.

It would be misleading, however, to conclude that the role of the 3D effects in the formation of the barium lines in the atmospheres of red giants in NGC~6752 is minor. As a matter of fact, we have found that the 3D--1D abundance corrections owing to horizontal temperature inhomogeneities in the 3D model (i.e., $\Delta_{\rm 3D-\langle3D\rangle}$ correction) \textit{and} differences in the temperature profiles between the average $\langle{\rm 3D}\rangle$ and 1D models ($\Delta_{\rm \langle3D\rangle-1D}$ correction) are substantial and may reach $\sim\pm0.1$\,dex (Table~\ref{tab:3D_Ba_Abn}). However, their sign depends on the line strength, and owing to this subtle fine-tuning their sum is significantly smaller, from --0.03 to 0.02\,dex, which for this given set of atmospheric and atomic line parameters maintains the size of the 3D--1D abundance corrections at the level of the errors in the abundance determination.

\begin{acknowledgements}
{We warmly thank David Yong for his support and valuable discussions during re-analysis of his spectroscopic data of NGC~6752. We also thank the referee Ian C. Short for many useful recommendations, which helped to improve the paper significantly. This work was supported in part by grants from the bilateral Lithuanian -- Ukrainian program (M/39-2009), the Lithuanian Science Council (TAP-35/2010, TAP-52/2010, and MIP-101/2011) and from the SCOPES grant No. IZ73Z0-128180/1. PB and AK acknowledge support from the Scientific Council of the Observatoire de Paris that allowed exchange visits between Paris and Vilnius. The paper is based on observations made with the European Southern Observatory telescopes obtained from the ESO/ST-ECF Science Archive Facility.}
\end{acknowledgements}

\bibliographystyle{aa}

\begin{thebibliography}{}


\bibitem[{Andrievsky} {et al.}(2009)]{ASK09}
Andrievsky,~S.~M., Spite,~M., Korotin,~S.~A., et al.
2009, \aap, 494, 1083

\bibitem[{Asplund}(2005)]{A05}
Asplund,~M.
2005, \araa, 43, 481

\bibitem[{Behara} {et al.}(2010)]{BBL10}
Behara,~N.~T., Bonifacio,~P., Ludwig,~H.-G., et al.
2010, \aap, 513, 72

\bibitem[{Bonifacio} {et al.}(2007)]{BPM07}
Bonifacio,~P., Pasquini,~L., Molaro,~P., et al.
2007, \aap, 470, 153

\bibitem[Caffau \& Ludwig (2007)]{CL07}
Caffau,~E., \& Ludwig,~H.-G.
2007, \aap, 467, L11

\bibitem[{Caffau} {et al.}(2011)]{CLS11}
Caffau,~E., Ludwig,~H.-G., Steffen,~M., Freytag,~B., \& Bonifacio,~P.
2011, \solphys, 268, 255

\bibitem[{Carlsson} (1986)]{carlsson}
Carlsson,~M.
1986, UppOR, 33

\bibitem[{Carretta} {et al.}(2009a)]{CBG09a}
Carretta,~E., Bragaglia,~A., Gratton,~R.~G., et al.
2009a, \aap, 505, 117

\bibitem[{Carretta} {et al.}(2009b)]{CBG09b}
Carretta,~E., Bragaglia,~A., Gratton,~R.~G., \& Lucatello,~S.
2009b, \aap, 505, 139

\bibitem[{Carretta} {at al.}(2010)]{CBG10a}
Carretta,~E., Bragaglia,~A., Gratton,~R.~G., et al.
2010a, \aap, 516, 55

\bibitem[{Carretta} {et al.}(2010b)]{CBG10b}
Carretta,~E., Bragaglia,~A., Gratton,~R.~G., et al.
2010b, \aap, 520, 95

\bibitem[{Castelli} (2005)]{C05}
Castelli,~F.
2005, \memsai, 8, 25

\bibitem[{Castelli} \& {Kurucz}(2003)]{CK03}
Castelli,~F., \& Kurucz,~R.~L.
2003, in: 'Modeling of Stellar Atmospheres', Proc. IAU Symp. 210, eds. N.E.~Piskunov, W.W.~Weiss, \& D.F.~Gray, poster A20 (CD-ROM); synthetic spectra available at http://cfaku5.cfa.harvard.edu/grids

\bibitem[{Cohen} (2011)]{C11}
Cohen,~J.
2011, \apj, 740L, 38

\bibitem[{Collet} {et al.}(2007)]{CAT07}
Collet,~R., Asplund,~M., \& Trampedach,~R.
2007, \aap, 469, 687

\bibitem[{Collet} {et al.}(2009)]{CAN09}
Collet,~R., Asplund,~M., \& Nissen,~P.~E.
2009, \pasa, 26, 330

\bibitem[Decressin et al.(2007)]{decressin}
Decressin,~T., Meynet,~G., Charbonnel,~C., Prantzos,~N., \& Ekstr{\"o}m,~S.
2007, \aap, 464, 1029

\bibitem[D'Ercole et al.(2011)]{dercole}
D'Ercole,~A., D'Antona,~F., \& Vesperini,~E.
2011, \mnras, 415, 1304

\bibitem[{Dobrovolskas} {et al.}(2010)]{DKL10}
Dobrovolskas,~V., Ku\v{c}inskas,~A., Ludwig,~H.-G., et al.
2010, Proc. of 11th Symposium on Nuclei in the Cosmos, Proceedings of Science, ID~288 ({\tt arXiv:1010.2507})

\bibitem[{Freytag} {et al.}(2012)]{FSL12}
Freytag,~B., Steffen,~M., Ludwig,~H.-G., et al.
2012, Journ. Comp. Phys., 231, 919

\bibitem[{Galazutdinov} (1992)]{G92}
Galazutdinov,~G.~A.
1992, Preprint SAO RAS., 92, 96

\bibitem[{Gonz\'{a}lez Hern\'{a}ndez} {et al.}(2009)]{GBC09}
Gonz\'{a}lez Hern\'{a}ndez,~J.I., Bonifacio,~P., Caffau,~E., et al.
2009, \aap, 505, 13

\bibitem[{Gratton} {et al.}(2001)]{GBB01}
Gratton,~R., Bonifacio,~P., Bragaglia,~A., et al.
2001, \aap, 369, 87

\bibitem[{Gratton} {et al.}(2004)]{GSC04}
Gratton,~R., Sneden,~C., \& Carretta,~E.
2004, \araa, 42, 385

\bibitem[{Harris} (1996)]{H96}
Harris,~W.~E.
1996, \aj, 112, 1487

\bibitem[{Ivanauskas} {et al.}(2010)]{IKL10}
Ivanauskas,~A., Ku\v{c}inskas,~A., Ludwig,~H.-G., \& Caffau,~E.
2010, Proc. of 11th Symposium on Nuclei in the Cosmos, Proceedings of Science, ID~290 ({\tt arXiv:1010.1722})

\bibitem[{James} {et al.}(2004)]{JFB04}
James,~G., Fran\c{c}ois,~P., Bonifacio,~P., et al.
2004, \aap, 414, 1071


\bibitem[{Korotin} {et al.}(1999)]{KAL99}
Korotin,~S.~A., Andrievsky,~S.~M., \& Luck,~R.~E.
1999, \aap, 351, 168

\bibitem[{Korotin} {et al.}(2011)]{KMG11}
Korotin,~S., Mishenina,~T., Gorbaneva,~T., \& Soubiran,~C.
2011, \mnras, 415, 2093

\bibitem[{Kraft}(1994)]{K94}
Kraft,~R.~P.
1994, \pasp, 106, 553

\bibitem[{Kupka} {et al.}(2000)]{KRP00}
Kupka,~F., Ryabchikova,~T.~A., Piskunov,~N.~E., Stempels,~H.~C., \& Weiss,~W.~W.
2000, BaltA, 9, 590

\bibitem[{Kurucz} {et al.}(1984)]{KFB84}
Kurucz,~R.~L., Furenlid,~I., Brault,~J., \& Testerman,~L.
1984, Solar flux atlas from 296 to 1300 nm.

\bibitem[{Kurucz}(1993)]{Kur93}
Kurucz,~R.~L.
1993, ATLAS9 Stellar Atmosphere Programs and 2 km/s Grid, CD-ROM No.13, Cambridge, Mass.

\bibitem[{Kurucz} (2005)]{Kur05}
Kurucz,~R.~L.
2005, \memsai, 8, 14

\bibitem[{Ludwig} {et al.}(2009)]{LCS09}
Ludwig,~H.-G., Caffau,~E., Steffen,~M., et al.
2009, \memsai, 80, 711

\bibitem[{Mashonkina} \& {Bikmaev}(1996)]{MB96}
Mashonkina,~L.~I., \& Bikmaev,~I.~F.
1996, ARep, 40, 94

\bibitem[{Mashonkina} {et al.}(1999)]{MGB99}
Mashonkina,~L., Gehren,~T., \& Bikmaev,~I.
1999, \aap, 343, 519

\bibitem[{Mashonkina} {et al.}(2011)]{MGS11}
Mashonkina,~L., Gehren,~T., Shi,~J.-R., Korn,~A.~J., \& Grupp,~F.
2011, \aap, 528, 87

\bibitem[{Mishenina} {et al.}(2009)]{MKA09}
Mishenina,~T.~V., Ku\v{c}inskas,~A., Andrievsky,~S.~M., et al.
2009, BA, 18, 193

\bibitem[{Norris} \& {Da Costa}(1995)]{NDC95}
Norris,~J.~E., \& Da Costa,~G.~S.
1995, \apj, 447, 680

\bibitem[{Norris} {et al.}(1996)]{NFM96}
Norris,~J.~E., Freeman,~K.~C., \& Mighell,~K.~J.
1996, \apj, 462, 241

\bibitem[{Otsuki} {et al.}(2006)]{OHA06}
Otsuki,~K., Honda,~S., Aoki,~W., Kajino,~T., \& Mathews,~G.~J.
2006, \apj, 641, L117

\bibitem[{Pasquini} {et al.}(2005)]{PBM05}
Pasquini,~L., Bonifacio,~P., Molaro,~P., et al.
2005, \aap, 441, 549

\bibitem[Piotto(2008)]{piotto08}
Piotto,~G.
2008, \memsai, 79, 334

\bibitem[Piotto(2009)]{piotto09}
Piotto,~G.
2009, in: 'The Ages of Stars', Proc. IAU Symp. 258, 233

\bibitem[{Ram\'{\i}rez} {et al.}(2009)]{RAK09}
Ram\'{\i}rez,~I., Allende Prieto,~C., Koesterke,~L., Lambert,~D.~L., \& Asplund,~M.
2009, \aap, 501, 1087

\bibitem[{Roederer} \& {Sneden}(2011)]{RS11}
Roederer,~I.~U., \& Sneden,~C.
2011, \aj, 142, 22

\bibitem[{Shen} {et al.}(2010)]{SBP10}
Shen,~Z.-X., Bonifacio,~P., Pasquini,~L., \& Zaggia,~S.
2010, \aap, 524L, 2

\bibitem[{Sbordone} {et al.}(2004)]{SBC04}
Sbordone, L., Bonifacio, P., Castelli, F., \& Kurucz, R.~L.
2004, \memsai, 5, 93

\bibitem[{Sbordone} (2005)]{S05}
Sbordone, L.
2005, \memsai, 8, 61

\bibitem[{Short} \& {Hauschildt}(2006)]{SH06}
Short,~C.~I., \& Hauschildt,~P.~H.
2006, \apj, 641, 494

\bibitem[{Sneden} {et al.}(1997)]{SKS97}
Sneden,~C., Kraft,~R.~P., Shetrone,~M.~D., et al.
1997, \aj, 114, 1964

\bibitem[{Sneden} {et al.}(2000)]{SPK00}
Sneden,~C., Pilachowski,~C.~A., \& Kraft,~R.~P.
2000, \aj, 120, 1351

\bibitem[Sneden et al.(2008)]{sneden}
Sneden,~C., Cowan, J.~J., \& Gallino,~R.
2008, \araa, 46, 241

\bibitem[{Sobeck} {et al.}(2011)]{SKS11}
Sobeck,~J.~S., Kraft,~R.~P., Sneden,~C., et al.
2011, \aj, 141, 175

\bibitem[{Suntzeff} \& {Kraft}(1996)]{SK96}
Suntzeff,~N.~B., \& Kraft,~R.~P.
1996, \aj, 111, 1913


\bibitem[{Wiese} \& {Martin}(1980)]{WM80}
Wiese,~W.~L., \& Martin,~G.~A.
1980, in Wavelengths and transition probabilities for atoms and atomic ions: Part 2. Transition probabilities, NSRDS-NBS Vol. 68

\bibitem[{Yong} {et al.}(2005)]{YGN05}
Yong,~D., Grundahl,~F., Nissen,~P.~E., Jensen,~H.~R., \& Lambert,~D.~L.
2005, \aap, 438, 875


\end{thebibliography}

\Online

\begin{appendix} 

\section{Spectral lines}

\begin{table}[t]
\begin{center}
\caption{List of iron lines used for atmospheric parameter determination of the target stars.
\label{tab:FeLines}}
  \begin{tabular}{cccc}
  \hline
  \noalign{\smallskip}
  Species & $\lambda$, nm & $\chi$, eV & log \textit{gf} \\
  \hline\noalign{\smallskip}
  Fe I   &  585.22187 & 4.548 & -1.330 \\
  Fe I   &  586.23530 & 4.549 & -0.058 \\
  Fe I   &  590.56720 & 4.652 & -0.730 \\
  Fe I   &  591.62474 & 2.453 & -2.994 \\
  Fe I   &  592.77891 & 4.652 & -1.090 \\
  Fe I   &  593.01799 & 4.652 & -0.230 \\
  Fe I   &  593.46549 & 3.928 & -1.170 \\
  Fe I   &  595.27184 & 3.984 & -1.440 \\
  Fe I   &  597.67750 & 3.943 & -1.310 \\
  Fe I   &  602.40580 & 4.548 & -0.120 \\
  Fe I   &  602.70509 & 4.076 & -1.089 \\
  Fe I   &  605.60047 & 4.733 & -0.460 \\
  Fe I   &  606.54822 & 2.608 & -1.530 \\
  Fe I   &  607.84910 & 4.795 & -0.424 \\
  Fe I   &  608.27106 & 2.223 & -3.573 \\
  Fe I   &  609.66653 & 3.984 & -1.930 \\
  Fe I   &  612.79066 & 4.143 & -1.399 \\
  Fe I   &  613.66153 & 2.453 & -1.400 \\
  Fe I   &  613.69947 & 2.198 & -2.950 \\
  Fe I   &  613.76917 & 2.588 & -1.403 \\
  Fe I   &  615.16181 & 2.176 & -3.299 \\
  Fe I   &  615.77284 & 4.076 & -1.260 \\
  Fe I   &  617.33356 & 2.223 & -2.880 \\
  Fe I   &  618.02042 & 2.727 & -2.586 \\
  Fe I   &  618.79904 & 3.943 & -1.720 \\
  Fe I   &  619.15584 & 2.433 & -1.417 \\
  Fe I   &  620.03129 & 2.608 & -2.437 \\
  Fe I   &  621.34303 & 2.223 & -2.482 \\
  Fe I   &  621.92810 & 2.198 & -2.433 \\
  Fe I   &  623.07230 & 2.559 & -1.281 \\
  Fe I   &  623.26412 & 3.654 & -1.223 \\
  Fe I   &  624.06462 & 2.223 & -3.233 \\
  Fe I   &  624.63188 & 3.602 & -0.733 \\
  Fe I   &  625.25554 & 2.404 & -1.687 \\
  Fe I   &  626.51340 & 2.176 & -2.550 \\
  Fe I   &  627.02250 & 2.858 & -2.464 \\
  Fe I   &  630.15012 & 3.654 & -0.718 \\
  Fe I   &  632.26855 & 2.588 & -2.426 \\
  Fe I   &  633.53308 & 2.198 & -2.177 \\
  Fe I   &  633.68243 & 3.686 & -0.856 \\
  Fe I   &  634.41491 & 2.433 & -2.923 \\
  Fe I   &  635.50290 & 2.845 & -2.350 \\
  Fe I   &  638.07433 & 4.186 & -1.376 \\
  Fe I   &  639.36013 & 2.433 & -1.432 \\
  Fe I   &  640.00012 & 3.602 & -0.290 \\
  Fe I   &  641.16493 & 3.654 & -0.595 \\
  Fe I   &  641.99496 & 4.733 & -0.240 \\
  Fe I   &  642.13508 & 2.279 & -2.027 \\
  Fe I   &  643.08464 & 2.176 & -2.006 \\
  Fe I   &  647.56244 & 2.559 & -2.942 \\
  Fe I   &  648.18703 & 2.279 & -2.984 \\
  Fe I   &  649.49805 & 2.404 & -1.273 \\
  Fe I   &  649.64666 & 4.795 & -0.570 \\
  Fe I   &  651.83671 & 2.831 & -2.460 \\
  Fe I   &  659.38705 & 2.433 & -2.422 \\
  Fe I   &  660.91103 & 2.559 & -2.692 \\
  Fe I   &  663.37497 & 4.558 & -0.799 \\
  Fe I   &  667.79870 & 2.692 & -1.418 \\
  Fe I   &  670.35674 & 2.758 & -3.160 \\
  Fe I   &  675.01525 & 2.424 & -2.621 \\
  Fe I   &  680.68449 & 2.727 & -3.210 \\
  Fe II  &  599.13760 & 3.153 & -3.540 \\
  Fe II  &  608.41110 & 3.199 & -3.780 \\
  Fe II  &  614.92580 & 3.889 & -2.720 \\
  Fe II  &  624.75570 & 3.892 & -2.310 \\
  Fe II  &  636.94620 & 2.891 & -4.160 \\
  Fe II  &  641.69190 & 3.892 & -2.650 \\
  Fe II  &  643.26800 & 2.891 & -3.520 \\
  Fe II  &  645.63830 & 3.903 & -2.100 \\
  Fe II  &  651.60800 & 2.891 & -3.320 \\
  \hline
  \end{tabular}
\end{center}
\end{table}

\end{appendix}

\end{document}